\documentstyle[12pt]{article}

\newcommand{\EQ}{\begin{equation}}
\newcommand{\EN}{\end{equation}}
\begin{document}
\topmargin 0pt
\oddsidemargin=-0.4truecm
\evensidemargin=-0.4truecm
\renewcommand{\thefootnote}{\fnsymbol{footnote}}
\newpage
\setcounter{page}{1}
\begin{titlepage}
\begin{flushright}
IC/92/216\\
SISSA-141/92/EP\\
LMU-10/92\\
August 1992
\end{flushright}
\vspace*{-0.2cm}
\begin{center}
{\large 17 keV NEUTRINO  AND LARGE MAGNETIC MOMENT SOLUTION OF THE SOLAR
NEUTRINO PUZZLE}
\vspace{0.4cm}

{\large Eugeni Kh. Akhmedov${}^{(a,b,c)}$
\footnote{E-mail: akhmedov@tsmi19.sissa.it, ~akhm@jbivn.kiae.su},
Zurab G. Berezhiani${}^{(d,e)}$
\footnote{Alexander von Humboldt Fellow}
\footnote{E-mail: zurab@hep.physik.uni-muenchen.de, ~vaxfe::berezhiani},\\
Goran Senjanovi\'{c}${}^{(a)}$
\footnote{E-mail: goran@itsictp.bitnet, ~vxicp1::gorans},
Zhijian Tao${}^{(a)}$}\\
\vspace*{0.6truecm}
${}^{(a)}$\em{International Centre for Theoretical Physics,
I-34100 Trieste, Italy\\
${}^{(b)}$Scuola Internazionale Superiore di Studi Avanzati,
I-34014 Trieste, Italy\\
${}^{(c)}$Kurchatov Institute of Atomic Energy, Moscow 123182, Russia\\
${}^{(d)}$Sektion Physik der Universit\"{a}t M\"{u}nchen, D-8000 Munich-2,
Germany\\
${}^{(e)}$Institute of Physics, Georgian Academy of Sciences, Tbilisi
380077, Georgia\\}
\end{center}
\vspace*{0.2truecm}
\begin{abstract}
Zee-type models with Majorons naturally incorporate the 17 keV neutrino but
in their minimal version fail to simultaneously solve the solar neutrino
puzzle. If there is a sterile neutrino state, we find a
particularly simple solution to the solar neutrino problem, which besides
$\nu_{17}$ predicts a light Zeldovich-Konopinski-Mahmoud neutrino
$\nu_{light}=\nu_e+\nu_{\mu}^c$ with a magnetic moment being easily as large
as $10^{-11}\mu_B$ through the Barr-Freire-Zee mechanism.
\end{abstract}
\vspace{1.cm}
\end{titlepage}
\renewcommand{\thefootnote}{\arabic{footnote}}
\setcounter{footnote}{0}
\newpage
{\em A. Introduction}

Whether or not a 17 keV neutrino, coupled to $\nu_e$ through the mixing angle
$\theta_S \simeq 0.1$, actually exists is very much an open question
\cite{PRO,CONTRA}. Its existence would imply fascinating modifications of the
standard model, in particular it would strongly encourage the idea of
Majorons, the Goldstone bosons of a spontaneously broken lepton flavor or
lepton number symmetry. Namely, Majorons provide fast decay of $\nu_{17}$
needed to satisfy the cosmological bounds on the stability of such a heavy
neutrino. A particularly elegant and simple implementation of the Majoron
picture in the context of a Zee-type model \cite{Zee}, based on maximal abelian
flavor symmetry $U(1)_e\times U(1)_{\mu}\times U(1)_{\tau}$, was proposed
in ref. \cite{BH}. This symmetry is expected to be
spontaneously broken down (at a scale close to $M_W$) to the generalized ZKM
\cite{ZKM} lepton number symmetry $L_e-L_{\nu}+L_{\tau}$ \cite{V} needed to
bring the existence of $\nu_{17}$ in accord with all the phenomenological and
cosmological constraints.

Unfortunately, as nice as it is, the above picture cannot be of help to the
solar neutrino puzzle (SNP) \cite{SNP}, since the particle spectrum contains
$\nu_{17}\simeq \nu_{\tau}+\nu_{\mu}^c$ and a massless $\nu_e$, mixed through
$\theta_S$.
A new neutrino state must be postulated, if neutrino properties are the basis
of a solution to the SNP. The natural choice of an additional active neutrino
is excluded due to the LEP limit on the $Z^0$ decay width. Thus one is led
to postulate a sterile neutrino (one or more), which must also remain light in
order to play a role in the 17 keV neutrino physics.

The simplest extension of the above program is to add a sterile
neutrino $n_R$, enlarging the symmetry group to $G=U(1)_e\times U(1)_{\mu}
\times U(1)_{\tau}\times U(1)_n$. Here we show how this scenario paves a way
for a simple and natural solution to the SNP based on the Barr-Freire-Zee
(BFZ) \cite{BFZ} mechanism of
generating a large magnetic moment $\mu_{\nu}$ of the neutrino. We imagine the
17 keV state to be $\nu_{17}\simeq \nu_{\tau}+n$ and a light ZKM state to be
$\nu_{light} \simeq \nu_e+\nu_{\mu}^c$, and show how $\nu_{light}$ can get a
large $\mu_{\nu}$ $(\sim 10^{-11}\mu_B)$. It is by now well known that the
large $\mu_{\nu}$ can flip the neutrino helicity in the magnetic field of the
sun, thus providing an alternative solution of the SNP (instead of neutrino
oscillations) \cite{C,VVO,ALM}. This could explain, if true, not only the
solar neutrino deficiency \cite{SNP}, but also the claimed anticorrelation of
the observed solar neutrino flux and the sunspot activity \cite{AC}.
The transition magnetic moments, which change neutrino helicity and
flavor simultaneously, can also do the job \cite{VVO,ALM}. In fact, it is the
transition moment between the components of a ZKM neutrino that we shall
utilize in this paper.

For a survey of a general situation regarding a role of $n$ in the 17 keV
neutrino physics, we refer the reader to our previous paper, in which both the
phenomenological and cosmological implications of this scenario were discussed
at length \cite{ABST}.

\vspace{0.4truecm}
{\em B. The model}

As we said in the introduction, we base our consideration on the $SU(2)_L\times
U(1)\times G$ symmetry, where $G=U(1)_e\times U(1)_{\mu}\times U(1)_{\tau}
\times  U(1)_n$. The model is a straightforward generalization of the one in
ref. \cite{BH}, which utilizes the lepton flavor symmetry in the framework of
the BFZ version of the Zee model. In other words, instead of one $SU(2)_{L}$
singlet charged scalar $h^-$, one introduces a set of such fields $h^-_{ab}$
$(a\not=b$, $a,b=e,\mu,\tau,n)$ which couple to leptons in the following
manner:
\begin{equation}
{\cal L}=f_{ij}l_i^TCi\tau_2l_jh_{ij}^*+f_{in}{e_{iR}}^TCn_Rh_{in}^*+h.c.
\end{equation}
Here $l_i$ are the usual lepton doublets, $e_{iR}$ are the singlet
right-handed charged leptons and
$f_{ij}=0$ for $i=j$ due to the antisymmetry of the $l^Tl$ terms.
Finally, to account for the breaking of lepton flavor symmetries and the
existence of Majorons, we introduce scalar $SU(2)_L\times U(1)$ singlets
$S_{ab}$ with the relevant couplings
\begin{equation}
\Delta V=\lambda_{ab}(\phi_1^Ti\tau_2\phi_2)h^{*}_{ab}S_{ab} +\lambda_{abcd}
h_{ab}h^{*}_{cd}S_{ab}^*S_{cd}+h.c.
\end{equation}
where $\phi_1$, $\phi_2$ are the $SU(2)_L\times U(1)$ Higgs doublets which
do not couple to fermions and are assumed to have zero VEVs (imagine a
symmetry $\phi_1\rightarrow {}-\phi_1$,  $\phi_2\rightarrow {}-\phi_2$, the
rest of the fields invariant). Furthermore, there is also the usual
Weinberg-Salam Higgs doublet $H$ which gives masses to charged fermions,
ensuring the natural flavor conservation.

The enlarged Higgs sector is actually minimal if one desires to have a BFZ
mechanism, natural flavor conservation and no electron neutrino mass at the
one-loop level. By giving up the last requirement, one could utilize a
simpler version of the model with only two scalar doublets $\phi_u$ and
$\phi_d$ separately coupled to up and down fermions. We shall comment on
this possibility in the {\em Discussion}.

{}From (1) and (2), the quantum numbers of $h$ and $S$ fields under $G$ are
($S$ and $h$ transform in the same manner)
\begin{equation}
L_ah_{bc}=(\delta_{ab}+\delta_{ac})h_{bc}
\end{equation}
where $L_a$ $(a=e,~\mu,~\tau,~n)$ are the lepton flavor charges.
In what follows we assume the symmetry breaking pattern $<S_{e\mu}>\not=0\not=
<S_{\mu\tau}>$, $<S_{\mu n}>\not=0$ which corresponds to the $G$
spontaneously broken down to ${\hat L}=L_e-L_{\mu}+L_{\tau}+L_n$. The neutrino
mass matrix which follows from our choice of $\nu_{17}$ and
$\nu_{light}$ is then [$(n^c)_L=C{\bar n}^T_R$]
\begin{equation}
\begin{array}{cc}
 & {\begin{array}{cc} \nu_{\mu} & n^c\end{array}}\\{}
M_{\nu}~=~\begin{array}{c}
\nu_e\\{} \nu_{\tau}\end{array}&{\left(\begin{array}{cc}
a&m\\{}b&M\end{array}\right)}\end{array}
\end{equation}
The elements $M$ and $m$ are generated at the one-loop level through the
diagrams shown in Fig. 1, whereas $a$ and $b$ appear only at the two-loop
level. We come to their values below when we discuss the relevant magnetic
moments.

The eigenvalues of the above matrix are
$$m_{1,2}\simeq \pm (a\cos\theta-b\sin\theta)$$
\begin{equation}
m_{3,4}\simeq \pm \sqrt{M^2+m^2}
\end{equation}
indicating two four-component states, with the Simpson mixing angle
$\theta_S\simeq \theta=\tan^{-1}(m/M)$, and we assume $M\simeq 17$ keV
and $a,b\ll m$.
{}From (1)-(3), the large elements of the mass matrix are (see Fig. 1)
$$M\simeq\displaystyle{\frac{1}{16\pi^2}f_{\mu\tau}f_{\mu n}
\lambda_{\tau\mu\mu n} m_{\mu}}$$
\EQ
m\simeq\displaystyle{\frac{1}{16\pi^2}f_{e\tau}f_{\tau n}
\lambda_{e\tau\tau n} m_{\tau}}
\EN
where we take all the scalar masses and the non-vanishing VEVs to be at the
electroweak scale\footnote{In any case, the difference in the masses can
be reabsorbed in the unknown coupling constants}.

Although in eq. (6) $M$ can be naturally of the order of 10 keV, the mixing
angle cannot be predicted. Moreover, $\theta\sim 0.1$ requires adjusting
a ratio of the
parameters by two orders of magnitude, since $m\propto m_{\tau}$.

In short, we have a Dirac 17 keV neutrino and a ZKM light neutrino.

\vspace{0.4truecm}
{\em C. Light neutrino: its magnetic moment and mass}

If the lepton charge ${\hat L}$ is left unbroken, we cannot have
neutrino oscillations as the solution to the SNP. It is conceivable, however,
that this breaking could come from the tiny gravitational effects, an idea
that has been discussed in \cite{HDO,ABST}, the one which we choose not
to pursue here.

In this paper we rather investigate the possibility that it is the neutrino
spin flip in the magnetic field in the sun that does the job. Our motivation
is twofold: first, we wish to stress that no breaking of ${\hat L}$ is
necessary and second, the beautiful BFZ mechanism of generating the large
magnetic moment for $\nu_{light}$ is naturally operative in our scenario.

Let us now address the BFZ mechanism \cite{BFZ}. The essential ingredient
is the two loop diagram of Fig. 2 which gives the transition neutrino magnetic
moment through the $h-W-\gamma$ coupling, $h$ being the Zee scalar (in
our case $h_{ab}$), and $\gamma$ denoting the photon. The point is
that the same diagram without the photon, i.e. the diagram that leads to the
neutrino mass, must
involve the longitudinal $W$ (or the unphysical Higgs in the $R$ gauge) and
so must be proportional to $m_l^2$, $m_l$ being the mass of charged lepton in
the loop. This is a remarkable result, since it provides a natural source of
a large magnetic moment, while keeping the neutrino mass small. Taking  as
before all the scalar mass parameters of the order of the electroweak scale,
one estimates the $\nu_e-\nu_{\mu}$ transition moment
\begin{equation}
\mu_{\nu}\simeq \left(\frac{1}{16\pi^2}\right )^2f_{e\mu}\lambda_{e\mu}
\lambda_{12}e\,g^{2}/M_W
\end{equation}
where $g$ is the electroweak gauge coupling constant, $\lambda_{12}$ is
the $\phi_1^{\dagger}\phi_2 H^{\dagger}H$ coupling constant, and so
\begin{equation}
\mu\leq 10^{-9}f_{e\mu}\lambda_{e\mu}\lambda_{12}\mu_B\leq 10^{-11}\mu_B
\end{equation}
since $f_{e\mu}{}^<_\sim 1/20$ from the universality of muon decay \cite{BP}.
We should stress that in our case all the dimensional scalar couplings of the
Zee model become VEVs of the $S$ fields.

The same diagram of Fig. 2 with the photon line removed gives the light
neutrino mass, i.e the mass matrix elements $a$ and $b$. Clearly they are
very small, of the order of $10^{-3}-10^{-1}$ eV \cite{BFZ}.

\vspace{0.4truecm}
{\em D. Discussion}

The central result of our paper is that one can reconcile the existence of the
17 keV neutrino with the solution of SNP through the transition magnetic
moment between the electron and muon neutrinos. The price for this is the
introduction of at least one sterile neutrino which is supposed to be
a part of Dirac $\nu_{17}$ state. The nice feature of this scenario is that,
contrary to the neutrino oscillation solution to the SNP, the generalized ZKM
symmetry ${\hat L}=L_e-L_{\mu}+L_{\tau}+L_{n}$ need not be broken.

Although perfectly consistent with all the laboratory data, this scenario
could at first glance run into difficulties with the cosmological and
astrophysical constraints on massive Dirac neutrinos. The astrophysical
constraints originate from the observation of the neutrino
signal from supernova 1987A and the fact that the assumed Dirac nature of
$\nu_{17}$ could give rise to a copious production of its sterile component in
the core of supernova. This comes about due to the helicity-flip scattering of
active neutrinos in the matter. The produced sterile neutrinos could shorten
the
neutrino pulse from supernova \cite{SNB1} or, upon decay, create a flux of
energetic $\nu_e$ or $\nu_{\mu}$ \cite{SNB2} in contradiction with
observations. The first point yields a constraint $m_D\leq 10-50$ keV which
does not in fact rule out the Dirac $\nu_{17}$; the second one can be evaded,
e.g. if one takes into account the possible reflip of right-handed sterile
neutrino into an active state in the supernova core due to resonant spin
precession\footnote{This mechanism was considered in connection with the
supernova constraints on the magnetic moments of Dirac neutrinos by Voloshin
\cite{Vol}.}.

The cosmological
constraint comes from the possibility of having sterile neutrino in
equilibrium at the nucleosynthesis epoch through the neutrino oscillations
\cite{CL1}. This would  be in apparent contradiction with the limit on
the number of neutrino species $N_{\nu}\leq 3.4$ as claimed in \cite{Chic}.
However, this may only happen in our model if the generalized ZKM lepton
charge ${\hat L}$ is broken and $\nu_{17}$ is a pseudo-Dirac particle, which
we do not really need.

We would like to offer here a few comments on the nature of the neutrino
transition due to $\nu_e-\nu_{\mu}$ magnetic moment in our model. If the
generalized ZKM symmetry is exact, the neutrinos undergo non-resonant
spin-flavor precession $\nu_e\rightarrow \bar{\nu}_{\mu}$, unless the magnetic
field is rotating along the neutrino trajectory \cite{SAKS}. In either
case, the neutrino transition
probability is energy independent, which could make it hard for this scenario
to reconcile Homestake and Kamiokande data with the recent GALLEX results
\cite{GALLEX}. Such a reconciliation requires energy dependence of the
transition probability.
This can be achieved easily if the generalized ZKM lepton charge is somehow
broken, which induces mass splits between the components of the light and
17 keV neutrinos. The resonant spin-flavor precession $\nu_e \rightarrow
{\bar \nu}_{\mu}$ \cite{ALM} which occurs in this case has the necessary
energy dependence discussed above\footnote{The $\bar{\nu}_e$ overproduction
problem which might arise in this case \cite{nubar} can be evaded in a number
of ways, see \cite{APS}.}. The natural source of this breaking can be e.g.
quantum gravity effects suggesting $\Delta m^<_\sim 10^{-5}$ eV
\cite{HDO,ABST}; the relevant quantity $\Delta m^2$ depends then on the
neutrino mass, i.e. the parameters $a$ and $b$. For $m_{\nu}{}^<_\sim 10$ eV
this gives $\Delta m^{2}{}^<_\sim 10^{-4}$ eV${}^2$, which is exactly
in the required range for the resonant spin-flavor precession scenario.
For the BFZ values $m_{\nu}\simeq m_{\mu}^2\,\mu_{\nu}\simeq (10^{-2}-
10^{-1})$ eV, one predicts $\Delta m^{2}_{light}\simeq (10^{-8}-10^{-6})$
eV${}^2$ which is still large enough to allow for the resonant effect.

As we noted above, for a pseudo-Dirac $\nu_{17}$ one may get in conflict
with the cosmological constraint $N_{\nu}\le 3.4$. However, there is no
universal agreement on the allowed value of $N_{\nu}$ and even $N_{\nu}=5$
was found acceptable by the authors of ref. \cite{Sar}.

Another way of evading this limit is the possible decay of $\nu_{17}$ into
$\nu_e$ with a lifetime of the order $\tau_{17}\simeq 10^{-2}$ s or so; the
produced excess of $\nu_e$'s can then compensate for the extra neutrino
species \cite{CL3}. In this kind of models, based on the extension
of the idea of flavons \cite{BH}, this is naturally achieved through decay
$\nu_{17}\rightarrow\nu_{light}+$ flavon.

We comment now on the simple version of the BFZ mechanism with natural
flavor conservation, i.e. the one with only two doublets $\phi_u$ and
$\phi_d$, coupled separately to up and down fermions \cite{ST}. This can be
ensured by a discrete symmetry: $\phi_{u} \rightarrow -\phi_{u}$,
$u_{R}\rightarrow -u_{R}$, $S\rightarrow -S$ with all the other fields
invariant. The eventual spontaneous symmetry breaking of the above symmetry
leads to the existence of domain walls, but it can be shown \cite{DW} that
instantons cause their decay thus lifting this serious cosmological problem.
In this case both $<\phi_u>\not=0\not=<\phi_d>$, allowing for the $h-W-
\gamma$ mixing. One obtains $\mu_{\nu}$ in the same manner as in the
original version or the one of Babu {\em et al.} \cite{Babu} of the BFZ
mechanism.

The problem is the one-loop generation of the neutrino mass matrix elements
$a$ and $b$. Since now only $\phi_d$ couples to charged leptons, its
couplings are $g_l \sim m_l/<\phi_d>\sim m_l/M_W$, $m_l$ being the charged
lepton mass. In the same manner as in the case of $M$ and $m$, one can
estimate
$$a\simeq\displaystyle{\frac{1}{16\pi^2}f_{e\mu}g_{\mu}
\lambda_{e\mu} m_{\mu}}$$
\begin{equation}
b\simeq\displaystyle{\frac{1}{16\pi^2}f_{\tau\mu}g_{\tau}
\lambda_{\tau\mu} m_{\tau}}
\end{equation}
By demanding $\mu_{\nu}\simeq 10^{-11}\mu_B$ we obtain $a\simeq 10$ eV, and
using $f_{\mu\tau}{}^<_\sim 10^{-1}$, $\lambda_{\mu\tau}{}^<_\sim 1$ we get
$b^<_\sim 10$ keV. The prediction for $a$ is very interesting since making
it smaller would decrease $\mu_{\nu}$, but unfortunately $b$ has to be
fine-tuned to make $\nu_e$ sufficiently light. Since it enters the
expression for the light neutrino mass being multiplied by $\theta_S$,
the fine tuning is not so bad; it is similar to the adjustment of the
coupling constants needed to make $m/M \simeq 0.1$. It is reasonable to
conclude that in this case the light neutrino mass should be very close
to its experimental upper limit $\sim 10$ eV.

\vspace{0.4cm}
{\em Acknowledgements}

We would like to thank K.S. Babu, K. Enqvist and K. Kainulainen for useful
discussions and Professor Abdus Salam, the International Atomic Energy Agency
and UNESCO for hospitality at the International Centre for Theoretical
Physics, Trieste. E.A. is grateful to SISSA for its kind hospitality.\\

\noindent
{\em Note added}. After this work had been completed we became aware of the
paper by Choudhury and Sarkar \cite{ChSa} who also study the issue of a
large neutrino magnetic moment in the 17 keV neutrino picture. Their model,
however, is completely different from ours.

\newpage
\centerline {\bf \large Figure captions}

\vglue 1truecm
\noindent
Fig. 1. One-loop diagrams responsible for the generation of a heavy neutrino
$\nu_{17}$ mass. All the fermion fields are left-handed.

\vglue 1truecm
\noindent
Fig. 2. Two-loop diagrams providing neutrino transition magnetic moment
$\mu_{\nu}$.
\end{document}